\documentclass[12pt]{article}
\usepackage[a4paper]{geometry}
\usepackage{calc}
\usepackage{color}
\usepackage{amsfonts}
\usepackage{latexsym}
\usepackage{placeins}
\ifx\pdftexversion\undefined
  \usepackage[dvips]{graphicx}
\else
  \usepackage[pdftex]{graphicx}
\fi
\usepackage{amssymb}
\usepackage{authblk}
\usepackage{amsmath}

\renewcommand\Authfont{\scshape\small}
\renewcommand\Affilfont{\itshape\small}
\newcommand{\smalllineskip}{\baselineskip=15pt}

\renewenvironment{abstract}[0]{\small\rm
        \begin{center}ABSTRACT
        \\ \vspace{8pt}
        \begin{minipage}{5.2in}\smalllineskip
        \hspace{1pc}}{\end{minipage}\end{center}\vspace{-1pt}}
\newcommand{\emailaddress}[1]{\newline{\sf#1}}

\let\LaTeXtitle\title
\renewcommand{\title}[1]{\LaTeXtitle{\large\textsf{\textbf{#1}}}}

\title{GMRT observations of the radio source $4C \;35.06$: precessing jets from a cD galaxy under assembly?}
\date{}

\author[1,6]{Biju ~K.G}
\author[2]{M.~Pandey-Pommier}
\author[3]{Sunilkumar~P}
\author[4]{Samir Dhurde}
\author[4]{Joydeep Bagchi}
\author[5]{Ishwara-Chandra~C.H}
\author[6]{Joe Jacob}
\affil[1]{WMO Arts \& Science college, Muttil, Kerala, India-673122
 \emailaddress{kgbiju42@gmail.com}}
\affil[2]{Centre de Recherche Astrophysique de Lyon- Observatoire de Lyon, 9 avenue Charles Andr\'e, \\Saint-Genis Laval, F-69230, France
}
\affil[3]{Govt. College, Malappuram, Kerala, India
}
\affil[4]{ Inter University Centre for Astronomy and Astrophysics(IUCAA), P.B No.7, Ganeshkhind, Pune-411 007, India
}
\affil[5]{ National Centre for Radio Astrophysics, TIFR, P.B No.3,Ganeshkhind, Pune-411 007, India
}
\affil[6]{ Newmann College, Thodupuzha, Kerala, India, 685584
}

\begin{document}
\maketitle

\begin{abstract}
We report  GMRT  observation of the strong radio source $4C\; 35.06$, an extended ( $z=0.047$) radio-loud AGN at the center of galaxy cluster $Abell\; 407$. The radio map at 610 MHz reveal a striking, helically twisted jet system emanating from an optically faint AGN host. The radio morhology closely resembles the precessing jets of the galactic  microquasar $SS\;433$. The optical SDSS images of central region show a complex ensemble of  nine galactic condensations within  $1 arc \ minute $,  embedded in a faint, diffuse stellar halo. This system presents a unique case  for studying the  formation of a giant elliptical galaxy (cD) at the cluster center. 
\end{abstract}

\section{Introduction}\label{s:intro}
The formation of cD galaxies are suggested to be occurring  at the centre of rich galaxy clusters in which multiple merger of  galactic members take place. The central region of the galaxy cluster $Abell\; 407$, is the most convincing and rare example of such a merger in the offing (Schneider et al.$ 1982$).  This region, where a  conglomeration of nine  galaxies shrouded in a stellar halo  is a typical case for studying the processes involved in such  mergers. In this study we focus mainly on the morphology of the jet from the radio source $4C\;35.06$ hosted by the cluster $Abell \;407$.

%
\begin{figure}
\centering
\includegraphics[width=12cm]{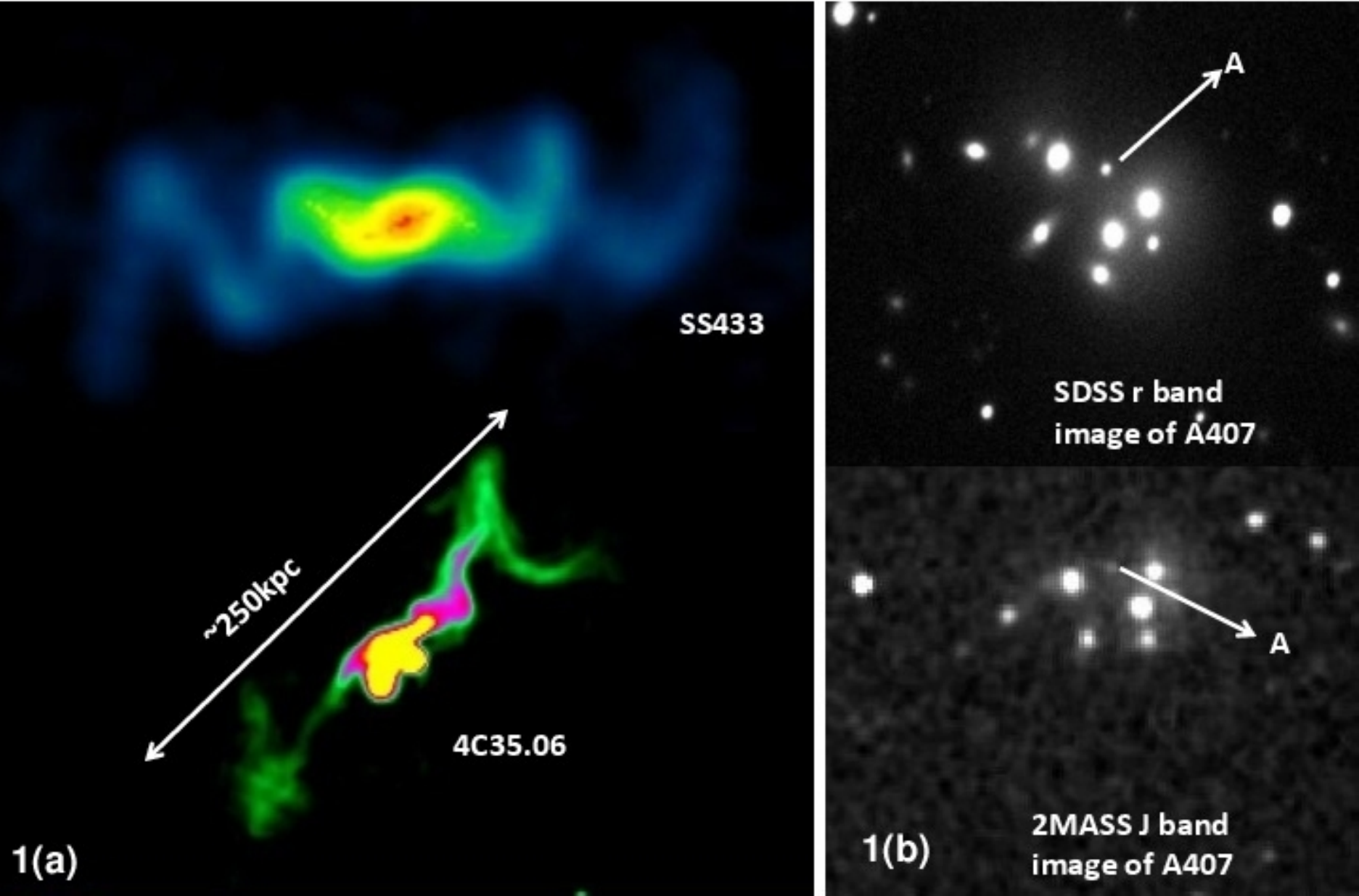}
\caption{(a) $SS \;433$ radio map(top)and 610~MHz radio map of $4C \;35.06$(bottom);(b)The SDSS r band and $2MASS$ J band images of the central region of $A \;407$. The galaxy, marked as A appears as the AGN host }  
\end{figure}
\section{Radio and optical studies of the source $4C35.06$}
The low frequency maps  from GMRT data show extended faint diffuse emission from the jet structure of $4C \;35.06$.  The  610~MHz radio map  shows a jet with helically twisted morphology and with a size estimate of approximately  $ 250\;kpc$, closely resembling that of the well known microquasar $SS \;433$ (Blundell and Bowler 2004).  It points to the fact that the jet is precessing and that the basic processes in jet emission in the stellar scales is the same in galactic scales also(Fig-1(a)). Also there are no  hot spots observed in the jet, indicating the continuous shifting of jet  direction. \\ The SDSS and 2MASS images show a faint galaxy coincident with the central region of the radio map with a positional accuracy of 1.5\textquotedblright \ (Fig-1(b)). Since powerful radio jets are observed to emanate only from massive elliptical galaxies, this observation seems to be  very unusual. For this low luminous optical host, possibly majority of its stars might have been stripped off during the violent mergers, still retaining a massive black hole at the centre. 
\\
\textbf{Results and conclusion}:The radio source $4C \;35.06$ hosts  a precessing  kpc scale jet  similar to that of the microquasar $SS \; 433$, and is seen to be emanating from the faintest galactic  member.  $4C \;35.06$ underscores the paradigm that despite the vast difference in  scales, jet phenomena in  radio loud AGNs keep a striking similarity with that in microquasars.

\end{document}